\documentclass[runningheads]{llncs}
\usepackage{graphicx}
\usepackage{booktabs}
\begin{document}
\title{Climate Action During COVID-19 Recovery and Beyond: A Twitter Text Mining Study}
\titlerunning{Climate Action During COVID-19 Recovery and Beyond}
%
\author{Mohammad S. Parsa\inst{1} \and
Lukasz Golab\inst{1}\and
S. Keshav\inst{2}}
%
%
\institute{University of Waterloo, Canada 
\email{\{msparsa,lgolab\}@uwaterloo.ca}\\ \and
University of Cambridge, U.K.
\email{sk818@cam.ac.uk}}
\maketitle              
\begin{abstract}
  The Coronavirus pandemic created a global crisis that prompted immediate large-scale action, including economic shutdowns and mobility restrictions.  These actions have had devastating effects on the economy, but some positive effects on the environment.  As the world recovers from the pandemic, we ask the following question: What is the public attitude towards climate action during COVID-19 recovery and beyond?  We answer this question by analyzing discussions on the Twitter social media platform.  We find that most discussions support climate action and point out lessons learned during pandemic response that can shape future climate policy, although skeptics continue to have a presence.  Additionally, concerns arise in the context of climate action during the pandemic, such as mitigating the risk of COVID-19 transmission on public transit.

\keywords{text mining \and social media \and climate change \and COVID-19}
\end{abstract}

\section{Introduction}
The COVID-19 pandemic has created a global crisis.  Controlling the spread of the virus required immediate large-scale action, including shutdowns and mobility restrictions.  While these actions have had negative effects on the economy, the corresponding reduction in carbon emissions resulted in some positive impacts on the environment such as improvements in air quality \cite{Berman2020} and an increase in wildlife breeding success \cite{Manenti2020}.

As the world recovers from the pandemic, these positive environmental impacts are at risk of vanishing.  It has therefore been suggested that COVID-19 recovery programs should include climate action such as investing in sustainable infrastructure and technologies \cite{Cohen2020,Rosenbloom2020}.  Others additionally suggest building on social changes such as working from home \cite{Belesova2020,Howarth2020}.

An important aspect of policymaking is an understanding of public opinion, especially now, given the social and economic sacrifices required to combat the pandemic.  We therefore ask the following question in this paper: \emph{What is the public attitude towards climate action during COVID-19 recovery and beyond?}  

Online social media platforms such as Twitter have been identified as critical tools for reflecting and predicting public opinion on a variety of topics.  We therefore answer our question by analyzing messages posted on Twitter during the first wave of the COVID-19 crisis (January to August 2020) that include keywords related to both the pandemic and climate change.  

Our methodology consists of the following steps.  First, we apply a \emph{topic modelling}  algorithm that segments the tweets based on the words used in order to identify topics of discussion.  Next, we measure the sentiment of opinions expressed on each topic as well as the percentage of tweets classified as inflammatory.  Finally, we inspect a sample of tweets belonging to each topic to confirm the nature of the topic and the sentiment of the opinions on this topic. To the best of our knowledge, our work is the first data-driven study of public commentary on climate action during the pandemic, as reflected on social media.  

The remainder of this paper is organized as follows.  Section~\ref{sec:related} discusses related work; Section~\ref{sec:methods} explains the data and methods used; Section~\ref{sec:results} discusses the results; and, Section~\ref{sec:discussions} concludes with insights and directions for future work.

\section{Related Work}
\label{sec:related}

There has been previous social media mining work (mainly using Twitter) on understanding attitudes towards climate science and climate action; see \cite{pearce2019social} for a recent survey.  The focus has been on discussion topics, spatiotemporal patterns, communities of activists and skeptics, correlations between Twitter activity, weather and political events, political polarization, public engagement in climate discussions, as well as scientific consensus on climate change.  However, these studies were done before the COVID-19 pandemic.

COVID-19 discussions on Twitter and other social media have also been studied recently.  Areas of focus include topic modelling \cite{Xue2020}, frequently asked questions \cite{anderson2017,Fernandez2016}, effects of the pandemic on mental well-being \cite{Gao2020}, and impacts of rumors and misinformation \cite{Apuke2021,Cinelli2020,Pennycook2020,Tasnim2020}.  However, climate-related discussions have not been studied in detail, and we fill this gap in this paper.

\section{Data and Methods}
\label{sec:methods}

This study uses data from the Twitter social networking and microblogging platform. Twitter users post messages with up to 280 characters, containing text, images, hyperlinks or hashtags, which are words starting with the symbol ‘\#’ and are used to index keywords and topics.  Users may follow, i.e., ask to receive tweets from, other users, and may forward, i.e., retweet, messages written by other users. There are currently over 330 million active users on Twitter worldwide, sending approximately 500 million tweets per day.

Recent work has identified over 81 million tweets containing words related to the COVID-19 pandemic, spanning from January 1, 2020 to July 31, 2020 \cite{Chen2020}. This dataset contains only the tweet IDs, and we obtained the full tweets via the public Twitter download interface.  This dataset only contains ‘public’ tweets and omits those marked as ‘private’ and therefore only visible to one’s followers.  

Following the methodology used in prior work on climate change discussions on Twitter \cite{Cody2015,Jang2015,Kirilenko2014}, we identified a subset of these 81 million tweets that contain at least one of the following terms: climate change, \#climatechange, global warming or \#globalwarming.  We then removed non-English tweets, leaving 155,716 tweets, and we removed retweeted copies, finally resulting in 39,461 tweets for analysis. While removing retweeted copies, we labelled each distinct tweet with the number of times it was retweeted.

Our analysis relies on topic modelling, which is a text mining tool that segments a collection of documents (in this case, tweets), with each segment containing documents that use similar words and thus are likely to describe the same topic.  To prepare the tweets for topic modelling, we performed standard text pre-processing, as in related work on social media mining \cite{Missier2016,Ostrowski2015}.  
We converted all letters to lower case, and we removed punctuation symbols, hyperlinks and stopwords (which are words that serve a grammatical purpose but do not convey any semantic meaning, such as ``and'', ``the'', etc.).  We then lemmatized the remaining words.  Lemmatization groups together the inflected forms of a word. For example, words such as ``plays'', ``played'' and ``playing'' are all lemmatized to ``play''.  We then vectorized each tweet based on the remaining words.  For each tweet, the $i$th entry of its vector corresponds to the Term Frequency-Inverse Document Frequency (TF-IDF) of the $i$th word in the set of (remaining) words occurring in the dataset. The TF-IDF score of a given word for a given tweet was computed by dividing the number of times the word appears in the tweet (TF) by the logarithm of the fraction of tweets that contain at least one occurrence of this word (DF).  TF-IDF is frequently used when vectorizing text since it considers both the uniqueness of a word in the dataset and the importance of the word to the specific document.

We applied the Non-negative Matrix Factorization (NMF) method for topic modelling \cite{xu2003} to the vectorized tweets.  NMF clusters the tweets into topics and produces a list of representative terms for each topic.  We report the following information for each topic.  

\begin{itemize}
    \item Each representative term is given a score by the NMF algorithm, and we selected the top-10 highest-scoring representative terms for each topic.
    \item We extracted the five most frequent word n-grams (for n up to three, i.e., sequences of up to three consecutive words) within the tweets assigned to each topic. 
    \item Following recent work on sentiment analysis of COVID-19 discussions on social media \cite{Sanders2020}, we calculated the percentage of tweets having a positive sentiment using the HuggingFace Transformer sentiment analyzer \cite{Wolf2019HuggingFacesTS}. 
    \item We calculated the percentage of comments identified as inflammatory using the HateSonar hate speech classifier \cite{DavidsonWMW17}, which was shown to work well on Twitter data. 
    \item To confirm the nature of the topics, we manually examined 100 most frequently retweeted tweets for each topic.
\end{itemize}

\begin{figure}[t]
  \centering
  \includegraphics[width=0.75\linewidth]{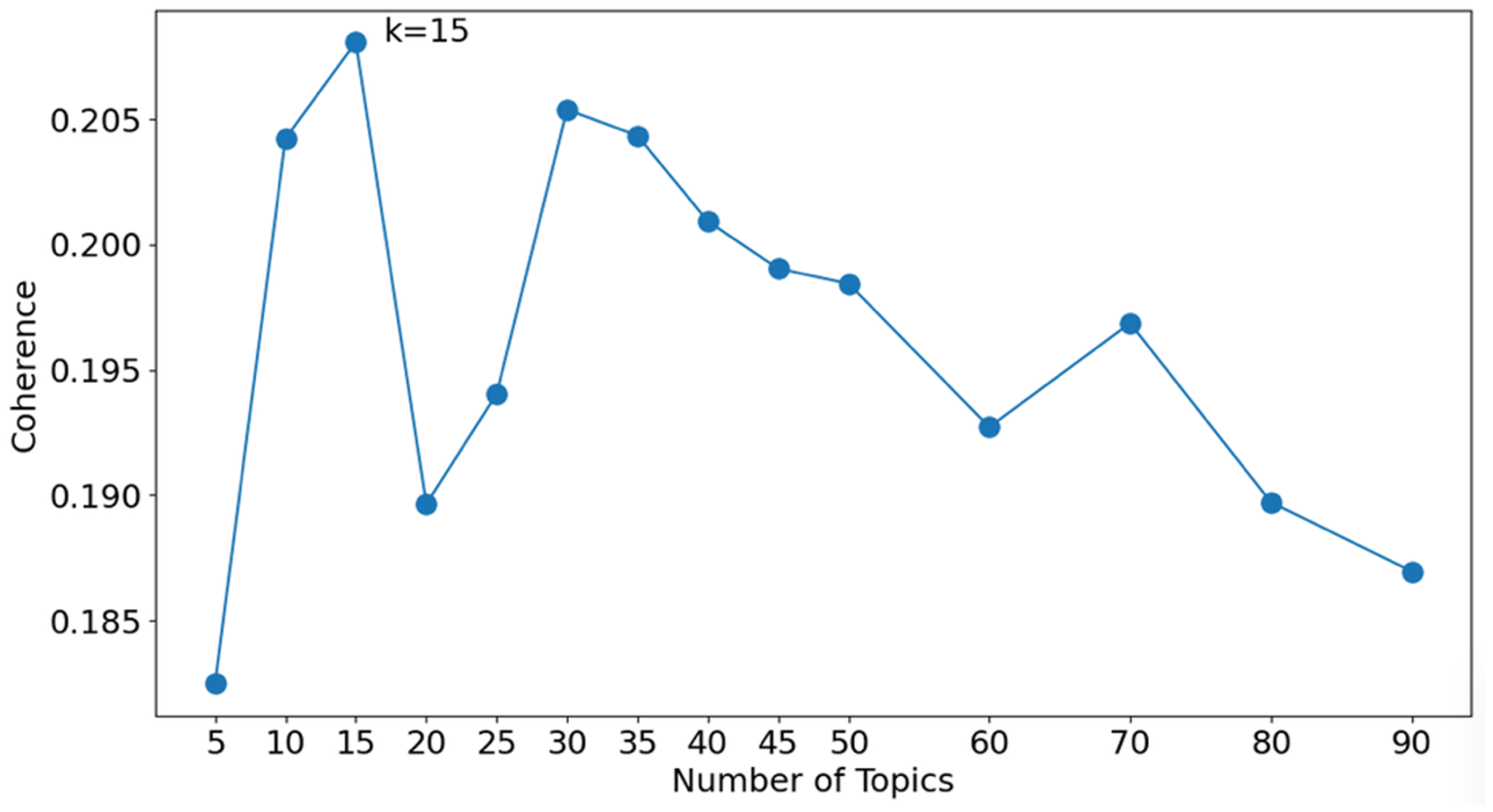}
  \caption{Coherence scores for different numbers of topics}
  \label{fig:coherence}
\end{figure}

NMF requires the number of topics as input. To select an appropriate number of topics, we ran NMF to produce between 5 to 90 topics and computed the coherence \cite{OCallaghan2015} of each output (higher is better). Coherence measures the extent to which the top representative terms representing each topic are semantically related.  As shown in Figure \ref{fig:coherence}, coherence was highest at 15 topics.

\section{Results}
\label{sec:results}

Topic modelling suggests two main themes in Twitter discussions: climate action during COVID-19 recovery (6 out of 15 topics; summarized in Table~\ref{tab:climate_action_topics}); and lessons learned for future climate action (8 out of 15 topics; summarized in Table~\ref{tab:lessons_learned_topics}).  These two themes cover all but one topic, containing specific discussions about China’s role in the pandemic and in the global carbon footprint.

Tables \ref{tab:climate_action_topics} and \ref{tab:lessons_learned_topics} contain the following information, from left to right: the topic number, the top-10 representative terms, the frequent n-grams, a topic description based on manual inspection of the most frequently retweeted tweets, the number of tweets assigned to the topic, the percentage of tweets with offensive language (OL), and the percentage of tweets with a positive sentiment (PS).  The topics are sorted by size, i.e., the number of tweets assigned to them.  In the remainder of this section, we discuss the two main themes in more detail.

\subsection{Theme 1: Climate action during COVID-19 recovery}

Table \ref{tab:climate_action_topics} summarizes the topics related to climate action during the pandemic and beyond.  Over 60 percent of the tweets in our dataset are related to this theme.  In terms of sentiment, topics 1, and 9 have the highest fraction of positive tweets, encouraging climate action during COVID-19 recovery.

\begin{table*}[t]
    \centering
    \caption{ Topics related to climate action during COVID-19 recovery}
  \label{tab:climate_action_topics}
    \begin{tabular}{cp{0.23\linewidth}p{0.22\linewidth}p{0.27\linewidth}ccc}
    \toprule
    &Representative terms & Frequent n-grams&Topic description& \# & \%OL& \%PS\\
    \midrule
    
1&need, time, new, threat, tackle, health, action, recovery, make, future& earthday 2020, public health ,  economic recovery,  recovery plan,  existential threat&Economic recovery plans should include climate action such as reducing fossil fuel use and CO2 emissions.&7901&1.5&31.4\\ \hline

2&human, pollution, emission, nature, earth, china, reduce, way, air, carbon&air pollution,  carbon emission,  fossil fuel,  human die, economic demographic&Economic shutdowns reduced CO2 emissions and air pollution, but more action is needed. However, activities such as using public transit become risky during the pandemic.&4793&1.7&20.7\\ \hline

7&think, thing, know, wait, bad, good, really, cause, hear, happen&people think,  think bad,  bad would,  think virus,  think serious&General discussions about the pandemic and long-term effects of climate change.&2468&4.8&19.4\\ \hline

9&fight, help, lesson, money, let, use, warren, elizabeth, deal, way&help fight,  fight back,  elizabeth warren,  deal chinese,  chinese must fight&Supportive and discouraging discussions around the use of pandemic recovery funds for climate action. &1932&1.8&31.8\\ \hline
 
12&stop, work, tax, pay, eat, animal, care, justice, racial, poor&stop eat,  eat animal,  racial justice,  poor racial justice,  stop eat animal&Suggesting a plan-based diet to protect the environment and prevent the spread of animal-borne diseases.&1423&6.1&18.6\\
 
\bottomrule
    \end{tabular}
\end{table*}

Topic 1 and 2, the most frequent topics, discuss recent improvements in air quality and express concerns about their temporary nature.  We show two example tweets below. 

\textit{``\#COVID19 locking down the whole world can be considered a LARGE SCALE EXPERIMENT for reduction of emissions. WE all can see the difference in the BLUE skies and through breathing clean air. Q: What will happen when \#Covid\_19 leaves us alone\#Enviroment \#ClimateChange'}

\textit{``COVID-19 shutdowns are clearing the air, but pollution will return as economies reopen — The shutdowns aren't slowing climate change.''}

As a result of these concerns, many tweets expressed the opinion that economic recovery plans should target climate action.  Suggested actions include reductions in the consumption of fossil fuels (Topic 1), encouraging a sustainable lifestyle with a plant-based diet (Topic 12), and investments in green transportation infrastructure such as protected cycle lanes and making public transit safe to use during the pandemic (Topic 9).  We give an example tweet below.

\textit{``Please support Clean State's campaign to develop a budget that supports sustainable employment and economic recovery package based on addressing a threat even greater than COVID-19: \#ClimateChange. Sign up to support our open letter.''}

In addition to calling on governments and world leaders to support climate action, many tweets suggested individual climate actions that can be done during the pandemic (Topic 1).  As an example, the \#DarkSelfieChallenge mentioned in one of the most frequently retweeted messages encouraged people to save energy by turning off their lights and taking a selfie in the dark:  

\textit{``Help the planet from home! With Hyundai, I'm taking part in the \#DarkSelfieChallenge. Turn all the lights off and take a selfie with the flash on. Show yourself in the dark to shed light on climate change. \#DarkSelfieChallenge \#EarthDay \#StayHome \#HyundaixBTS \#NEXO''}

Other examples include encouraging cycling as a sustainable mode of transportation that allows social distancing, as expressed in the following tweet.  

\textit{``All hail the mighty bicycle: France to pay for cycling lessons and bike repairs to fight both coronavirus and climate change - Bicycles promote physical distancing - Paris building 750km (466 miles) more bike lanes  - Bogota, Berlin, Brussels, Milan all going big on bicycles too''}

In contrast to the supporters of green recovery plans, some tweets argue that recovery plans should only target economic growth (topic 9), at least for now, and should suspend climate action for faster economic recovery.  According to these tweets, it would be more dangerous to the environment in the long run if the carbon tax drives businesses into bankruptcy, as this would lead to a weakened economy that is unable to fund climate action plans (Topic 7).  We show two example tweets below.

\textit{``The only way that would work is when the economies are stable and we doing it to tackle global warming and climate change. Where we indoors so planet earth can’t breathe. Otherwise we ain’t doing that **** as a remembrance''}

\textit{``Are you seriously that dumb to bring climate change and adding additional carbon tax at a time of this pandemic? The 75\% is for small to medium sized businesses. Rgat wont come free. They will pay double at the other end. Same for anyone deferring mortgages, car payments etc.''}

\subsection{Theme 2: Lessons learned during the pandemic to help combat climate change}

The second theme focuses on the lessons learned from the pandemic and discusses whether these lessons should be applied to mitigate climate change (Table \ref{tab:lessons_learned_topics}).  Topic 10 has a significantly high fraction of positive tweets that encourage applying lessons learned from the COVID-19 crisis to fighting climate change. Topics 11, 13 and 15, which criticize government policies and express skepticism towards the virus and climate change, have the lowest fraction of positive tweets. The fraction of offensive content is generally low, though it appears that skeptics express their thoughts more negatively and aggressively compared to other users

\begin{table*}[ht!]
    \centering
    \caption{Topics related to lessons learned during the pandemic to combat climate change}
  \label{tab:lessons_learned_topics}
    \begin{tabular}{cp{0.23\linewidth}p{0.22\linewidth}p{0.27\linewidth}ccc}
    \toprule
    &Representative terms & Frequent n-grams&Topic description& \# & \%OL& \%PS\\
    \midrule

3&say, science, scientist, believe, expert, listen, response, pope, deny, ignore&scientist say, believe science, listen expert, pope francis, anti science&Discussions of statements made by scientists, skeptics, and public figures.&3835&3.6&18.5\\ \hline
              
4&world, end, war, save, leader, react, happen, post, year, control& world war, world leader, end world, save world, world economy&The pandemic and climate change are issues that must be addressed by the whole world.&3047&3.4&26.8\\ \hline
        
5&people, die, care, believe, young, make, old, million, worry, imagine&people die, million people, old people, people believe, take seriously&The Coronavirus is dangerous for older adults; climate change is dangerous for future generations.&2718&5.7&17.7\\ \hline
 
8&like, look, disease, make, thing, feel, issue, treat, sound, right& disease like, issue like, infectious disease, threat like, disease threat&Climate change may become a crisis similar to or worse than the COVID-19 pandemic.&2340&5.6&21.4\\ \hline
 
10&crisis, warn, uk, avoid, lesson, opportunity, tackle, learn, climatecrisis, spur&health crisis, economic crisis, green recovery, avoid crisis, tackle crisis&Lessons learned from pandemic response should be applied to climate action.&1853&0.4&31.2\\ \hline

11&trump, blame, warren, elizabeth, disease, point, president, donald, long, obama&elizabeth warren, donald trump, blame trump, blame disease, blame disease like&Criticisms of the U.S. pandemic and climate policies.&1808&4.5&12.1\\ \hline

13&hoax, chinese, trump, real, impeachment, democrat, russian, fake, president, russia& call hoax, chinese hoax, virus hoax, impeachment hoax, hoax impeachment&Skepticism towards the Coronavirus and climate change, e.g., calling the Coronavirus a hoax.
&1059&5.6&10.1\\ \hline

14&kill, year, worry, million, end, heat, maybe, thing, right, want& kill people, kill million, virus kill, kill everyone, total death&Climate change may be as devastating and deadly as the Coronavirus crisis.&874&7.9&13.3\\ \hline

15&model, wrong, use, computer, accurate, predict, base, year, death, tell&computer model, model wrong, model predict, virus model&Skepticism towards pandemic and climate models.&802&1.2&13\\
\bottomrule
    \end{tabular}
\end{table*}

A frequently mentioned lesson (Topic 10) is that global consensus and collaboration were critical during the COVID-19 pandemic and will be critical in future climate action. It was observed that some countries implemented pandemic response measures early and were able to control the spread of the virus; similarly, preparation and preventive measures will be important in the context of climate action.  Additionally, specific pandemic response actions were mentioned as being helpful in the context of climate action, such as travel restrictions and working from home (Topic 10).  We give three example tweets below. 

\textit{``Great piece on the lessons that the Coronavirus response already has for climate change: early, effective and far-reaching action (dont wait for consequences to unfold) and international collaboration and solidarity.''}

\textit{``Inslee: ‘we should not be intimidated by people who say you should not use this COVID crisis to peddle a solution to climate change.’ He's using this crisis to push his agenda. He doesn't care about businesses, life savings, careers, lost.''}

\textit{``Global inaction on climate change offers grim lessons in the age of coronavirus: ‘Only in hindsight will we really understand what we gambled on’ one climate scientist said.' ''}

Another common opinion reflecting lessons learned was that climate change crisis may be similar to or worse than the Coronavirus, for example, in terms of fatality rates (Topic 14).  It was argued that while the virus is more dangerous for the older population, climate change will be dangerous for the next generation (Topic 5).  The following tweet is one of many that express this opinion. 

\textit{``Corona Virus is an existential threat to old people. Climate Change is an existential threat to young people. Recipe for a coalition of the many \#BernieSurge'}'

The third common discussion topic in the context of lessons learned was the role of science in decision and policymaking (Topic 3), as expressed by the following tweet.

\textit{``Diseases like coronavirus remind us why we need robust institutions and investments in public health, and a government that is ready to respond at any moment. That means using science-based policy and confronting climate change, which will affect how diseases emerge and spread.''}

On a related note, there were negative reactions to statements made by public figures who underestimated the seriousness of the Coronavirus and climate change (Topic 11).  However, Pope Francis’ statement that the ``Coronavirus pandemic could be nature's response to the climate crisis'' received a mixed response.  Some tweets agreed with this statement, while others pointed out a lack of scientific proof (Topic 3). 

On the other hand, a small minority of skeptics continued to doubt climate change and criticized any use of the pandemic to advance climate action (Topic 13), as shown in the following tweet.

\textit{``Ms Swedish environmental goddess her team of alarmed activists may try to link \#coronarovirus to \#ClimateChange  Won't be surprised if they made a speech relating \#ClimateEmergency to \#coronoavirus outbreak lol \#climatechangehoax \#ClimateHoax \#auspol \#auspol2020 \#Australia ''}

Furthermore, some tweets questioned the accuracy of climate and COVID-19 prediction models. They argued that these models are not reliable at estimating deaths caused by the virus and deaths that climate change will cause in the future (Topic 15).  We give an example tweet below.  

\textit{``Why would anyone trust the climate change models when the Covid-19 models have been off by an incredible (perhaps deliberate) amount?''}

 \section{Discussion and Conclusions}
\label{sec:discussions}

Our study of Twitter discussions on climate change during the COVID-19 pandemic revealed two main themes: Climate action during COVID-19 recovery and lessons learned for the future. 

Climate action during COVID-19 recovery is largely supported on Twitter, with positive opinions on actions such as investing in sustainability and education, as well as social changes such as promoting the use of bicycles and public transit.  
However, some tweets express concerns about the higher risk of contracting the Coronavirus when using public transit.  Thus, it appears that there is at least some willingness to continue making environmentally-friendly decisions during the pandemic as long as policies are in place to mitigate risks.  This finding should be of interest to local governments wishing to encourage climate-friendly behaviour during the pandemic.

Similarly, many tweets positively reflect on the lessons learned during the pandemic that may shape future climate action, such as the importance of preventive measures, the role of science in public policymaking, and the need for coordinated global action.  In particular, we identified a topic (Topic 8; Table~\ref{tab:lessons_learned_topics}) reflecting the opinion that climate change may become a worse global crisis than COVID-19, underscoring the need for global action.  Additionally, we found tweets that criticized politicians who ignore both climate change and COVID-19 safety protocols such as face masks, reinforcing the importance of making scientifically-sound decisions in the future (Topic 11; Table~\ref{tab:lessons_learned_topics}).  

On the other hand, as observed in pre-pandemic social media studies \cite{boussalis2016text,williams2015network,pearce2019social}, climate skeptics continue to have a presence on Twitter.  We additionally found that skeptics have incorporated the pandemic and the associated economic crisis into their reasoning for suspending climate policies such as the carbon tax.  Moreover, some tweets use skepticism for one issue to justify skepticism for another, as evidenced by tweets calling both climate change and the Coronavirus pandemic a hoax, and those that mistrust climate models because COVID-19 models are believed to be inaccurate.  An in-depth investigation on how misinformation about climate change creates misinformation about the pandemic and vice versa is an interesting direction for future research.

One limitation of this study is its focus on English language content on the Twitter platform.  Thus, another direction for future work is to compare public attitudes towards climate action during COVID-19 recovery in Asia, Europe and North America, and correlate these opinions with pandemic response policies. 

\bibliographystyle{splncs04}
\bibliography{samplepaper}
\end{document}